# BRAIN PROSTHESES AS A DYNAMIC SYSTEM
## *(IMMORTALIZING THE HUMAN BRAIN?)*

Vadim Astakhov, Tamara Astakhova
Center for Research in Biological Systems (CRBS) University of California San Diego
astakhov@ncmir.ucsd.edu, vadim_astakhov@hotmail.com

**Abstract**
Interest in development of brain prostheses, which might be proposed to recover mental functions lost due to neuron-degenerative disease or trauma, requires new methods in molecular engineering and nanotechnology to build artificial brain tissues. We develop a *Dynamic Core* model to analyze complexity of damaged biological neural network as well as transition and recovery of the system functionality due to changes in the system environment. We provide a method to model complexity of physical systems which might be proposed as an artificial tissue or prosthesis.
*Delocalization of Dynamic Core* model is developed to analyze migration of mental functions in dynamic bio-systems which undergo architecture transition induced by trauma. Term Dynamic Core is used to define a set of causally related functions and Delocalization is used to describe the process of migration. Information geometry and topological formalisms are proposed to analyze information processes. A holographic model is proposed to construct dynamic environment with self-poetic *Dynamic Core* which preserve functional properties under transition from one host to another. We found statistical constraints for complex systems which conserve a Dynamic Core under environment transition. Also we suggest those constraints might provide recommendations for nanotechnologies and tissue engineering used in development of an artificial brain tissue.

## Introduction

Recent attempts in linking the human brain to electronic devices and building brain prostheses [1] raise interesting questions about survival of human consciousness and how technologies can affect it. These questions cover the area of altered state of consciousness triggered by neuron-degenerative disease, trauma and organ transplantation. The topic of survival of the brain damage has been discussed in the philosophical domain through human history but in the last hundred years has become a subject of scientific research that emerges from medical attempts to restore mental functions lost due to a disease or a trauma. One of the frontiers in that science is the development of brain prostheses which emerges from pioneering research [2, 3] of Theodore Berger, the leader of a team of engineers and neuroscientists at the University of Southern California in Los Angeles.

Francis Crick [4] "the astonishing hypothesis" about our mind and consciousness consist entirely of physiological activity in the tissue of the brain lead to the assumption that complex mental functions emerged from complex brain neural architecture. Thus, recovery of brain functionality might require prosthesis media with a level of complexity equivalent to that of the unimpaired human brain. Rephrasing Searle: [4] the artificial heart do not have to be made of muscle tissue, but whatever physical substance they are made of should have a causal complexity at least equal to that of actual heart tissue where the term "causal complexity" reflects the quantity of causal relations. The same can be true for an artificial brain tissue that replicates functionality though it is made of something totally different then neurons as long as the *artificial part of the brain* shares the level of causal complexity found in natural brains. Figure 1 gives an illustration on hypothetical process of neuron-degeneration that lead to brain degradation. Also, it shows how an artificial prosthesis might recovery neural tissue degradation and as such recover functional properties

*Figure 1. Brain prosthesis restores lost cognitive function due to recovery of degradated anatomical features.*

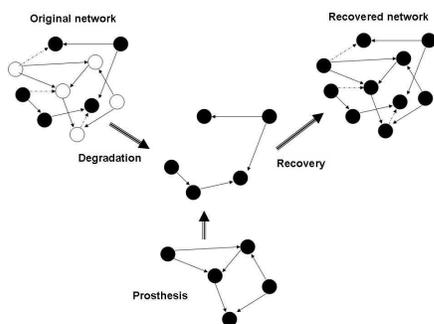

We propose new process called "*delocalization*" for recovery which is not as a direct copying of lost anatomical features but rather reestablishing lost functionality due to adjustment in the system: degraded brain system and artificial environment - prostheses. As an illustration, consider an example from network theory: where node A sends a message to the node C through the shortest path on the network. That case can be found in many biological systems. The "smart" node A should calculate the shortest paths among network nodes and send message to a proper neighbor. An arbitrary node should be able to perform a few basic operations which we schematically defined as: *get message*, *find next in the path*, and *send message to the next*. As can be easily seen, the system is not functioning if ability to find the shortest path destroyed. But likely, similar functionality can be recovered effectively if each node can be adjusted to perform just simple operations: *get message* and *send message to all neighbors*. The signal will be transmitted from A to C through many paths as well as through the shortest path. We can say that *find the shortest path* function is implicitly emerged in this model as opposed to explicitly implemented in the smart node model.

*Figure2. The left figure represents central-server architecture of communication nodes where a central general server A can sends a request to pass a message to node C through the pre-computed shortest path from A to C. The same functionality can be implemented (right) by broadcasting the message to all neighbors.*

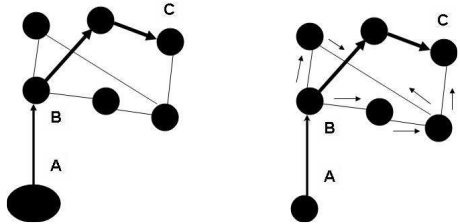

For example, the dynamic that will not broadcast the message but sent it to random node will not lead to emergence of the *shortest path* functionality. This is obviously very noisy example but it illustrates possibility to recover the lost functionality by modification in the system architecture.

More complex example provided in [Appendix A] where we consider *mediator* service [5] that integrate distributed data and provide mechanism of querying of those distributed sources.

## Delocalization as holography

The whole idea of *delocalization* seems analogous for well known effect of holographic representation which is well known from various parts of physics. Aside of optical applications there is a mathematical formalism that provide a basis for multiple representations of physical systems. A physical system can be described by holographic representation. That hologram is spatially distributed and requires none-local descriptors. Famous example from mathematical physics is duality between localized in the space physical systems in Anti-de Sitter (AdS) model and their holographic representation in Conformal Field Theory (CFT) [6]. Such holographic- delocalized representation preserve all causal relations of the localized system. It was proposed by Maldacena and demonstrated by Witten that certain conformal field theories in d dimensions can be described on the product of d+1 dimensional AdS space with a compact manifold.

We employ that approach for problem of re-implementation of lost functionality. First, we developed formalism to represent system functionality in geometrical terms. Geometrical formalism naturally merges with AdS formalism and CFT Yang-Mills theory used in work of Witten and others. Then, we suggest localized implementation of the functionality in d+1-dimmensional AdS model and delocalized holographic representation in d-dimensional CFT model. Finally, we consider a process of geometrical flow as a process of delocalization for pre-defined functionality. The process can be interpreted as a process of migration for system functionality from localized to delocalized representation.

Methods from the theory of *dynamical systems* are employed to provide geometrical formalism for analysis of network systems. A *dynamical system* is a mathematics concept in which a fixed rule describes the time dependence of a point in a geometric space. A *system state* is determined by a collection of numbers that can be measured. Small changes in the state of the system correspond to small changes in the numbers. We use the concept of *informational manifold* from "information and statistical geometry" [7] as the geometric *space* and introduce a *system state* as a point on the manifold to describe the dynamics of processes. The numbers are also the coordinates of a geometric space—a manifold. The *evolution* of the dynamical system is a rule that describes what and how future states follow from the current state.

# Information Geometry for Analysis of Dynamic Bio-System

We consider a biological network as a dynamic system X composed of n units $\{x_i\}$. Each unit can represent an object like a single neuron or a sub-net of the brain network. Those units can be either "on" or "off" with some probability. "On" means an element contributes to the activity that lead to emergence of the mental function Q and "off" is otherwise. Thus observable state of the function $Q = (Q_1, Q_2,..)$ for the system X can be characterized by certain sets of statistical parameters $(x_1, x_2, …)$ with given probability distribution $p(X|Q)$. The idea of endowing the space of such parameters with metric and geometrical structure leads to proposal of use Fisher information as a metric of geometric space for $p(X|Q)$-distributions:

$$g_{\mu\nu} = \int ( \partial p(X|Q) / \partial Q\mu) (\partial p(x|Q) / \partial Q\nu)) p(X|Q) d\{x_i\}.$$

Introduced Fisher metric is a Riemannian metric. Thus we can define distance among states as well as other invariant functional such as affine connection $\Gamma^{\sigma}_{\lambda\nu}$, curvature tensor $R^{\lambda}_{\mu\nu k}$, Ricci tensor $R^{\mu k}$ and Curvature scalar R [7] which describes the information manifold.

The importance of studying statistical structures as geometrical structures lies in the fact that geometric structures are invariant under coordinate transforms. These transforms can be interpreted as modifications of $\{x_i\}$ set by artificial tissue with different characteristics. Thus the problem of a system survival under transition from one bio-physical medium to another can be formulated geometrically.

*Evolution* of the network system can be modeled by Euler-Lagrange equations taken from small virtual fluctuation of metric for scalar invariants. One example of such evolution equation is

$$J = -1/16\pi \int \sqrt{det\ g^{\mu k}}\ (Q)\ R(Q)\ dQ.$$

But for open system such as neural network, external constraint can be added as a scalar term dependent on arbitrary covariant tensor $T^{\mu k}$:

$$J = -1/16\pi \int \sqrt{g(Q)}\ R(Q)\ dQ + 1/2 \int \sqrt{g(Q)}\ T^{\mu k}\ g_{\mu k}\ dQ.$$

That lead to well known geometrical equation $R^{\mu k}(Q) - g^{\mu k}(Q)R(Q) + 8\pi T^{\mu k}(Q) = 0$ which describe metric evolution. Solutions of this equation represent statistical systems under certain constraints defined by tensor $T^{\mu k}$. Thus, functionality of network systems can be resented in terms of gravitation theory.

Another way to employ geometrical approach is to define tangent vector "A" for each point X of manifold M(X) as: $A_\mu \sim \partial \ln(p(x))/\partial x_\mu$

Where Li brackets $[A_\mu A_\nu] \sim A_k$, give use the way to find transformations which will provide invariant descriptors. First we can employ approach developed in gauge theories that are usually discussed in the language of differential geometry that make it plausible to apply for informational geometry. Mathematically, a *gauge* is just a choice of a (local) section of some principle bundle. A gauge transformation is just a transformation between two such sections. Note that although gauge theory is mainly studied by physics, the idea of a connection is not essential or central to gauge theory in general. We can define a gauge connection on the principal bundle. If we choose a local basis of sections then we can represent covariant derivative by the connection form $A_\mu$, a Lie-algebra valued 1 –form which is called the gauge potential in physics. From this connection form we can construct the curvature form *F*, a Lia-algebra valued 2-form which is an intrinsic quantity, by $F_{\mu\nu} = \partial_\mu A_\nu - \partial_\nu A_\mu - ig[A_\mu\ A_\nu]$

$[D_\mu D_\nu] = -ig F_{\mu\nu}$, where $D_\mu = \partial_\mu - igA_\mu$; $A_\mu = A_\mu^a t^a$ and t –is generator of infinitesimal transformation

Thus we can write Lagrangian: $1/4\ F_{\mu\nu}^a F_{\mu\nu}^a \Leftrightarrow -1/2\ Sp(F_{\mu\nu} F_{\mu\nu})$ that is invariant under transformation of coordinates. Such approach provides us with informational analog of CFT Yang-Mills model.

At the same time, another evolution functional $J = \int (R + |\nabla f|^2) exp(-f) dV = \int (R + |\nabla f|^2) dm$, that is dependant on function f –gradient vector field defined on the manifold of volume V. It is well known in super-gravity (and string theory) [10] functional that can lead us to AdS model. Thus using statistical geometry approach, we can formulate network system functionality in terms of two models: super-gravity and Yang-Mills.

It can be easely shown that functional $J = \int (R + |\nabla f|^2) dm$ can be taken as the gradient flow $dg_{ij}/dt = 2(R_{ij} + \nabla_i \nabla_j f)$ that is generalization of the geometrical flow or so called Ricci flow $dg_{ij}/dt = -2R_{ij}$. Interesting thing about Ricci flow is that it can be characterized among all other evolution equations by infinitesimal behavior of the fundamental solutions of the conjugate heat equation. It is also related to the description of the renormalization group flow.

## Dynamic Core of a function

We introduce a definition *Dynamic Core* -- A dynamic system that consists of a set of dynamic elements causally interacting with each other and the environment in a way that lead to a high level of information integration within the system and emergence hierarchical causal interactions within the system. A higher level of causal power among system elements is compared to causal interactions with an environment. The dynamic core can be seen as a functional cluster characterized by strong mutual interaction among a set of sub-groups over a period of time. It is essential that this functional cluster be highly differentiated.

Dynamic Core (DC) will be used to describe any dynamical system that has a sub-part acting in causal relations with each other. To measure *causal relation* some metrics considered are *information integration* and *causal power*. Dynamic core is defined as a sub-system of any physical environment that has internal *information integration* and *causal power* [8] much higher than mutual *information integration* between the system and the environment.

To describe specialized sub-networks relevant to emergence of specific high level functions, we employ concept of functional cluster [9]: If there are any causal interactions within the system such as signals transfer, the number of states that the system can take will be less than the number of states that its separate elements can take. Some sub-nets can strongly interact within itself and much less with other regions of the brain. Geometrically, it is equivalent to higher positive curvature $R \sim CI(X^k_j) = I(X^k_j) / MI(X^k_j; X - X^k_j)$ [9] in the area of information manifold which reflects the system $X^k_j$ state dynamics due to the loss of information entropy "H". The loss is due to interactions among the system elements - $I(X^k_j) = \sum H(x_i) - H(X^k_j)$ and interaction with the rest of system described by mutual entropy $MI(X^k_j; X - X^k_j)$.

*Figure 3. System X partitioned to subset of elements $X^k_j$ and the rest of the system $X - X^k_j$. The dashed ellipse represents another possible partition. The dependence between the subset $X^k_j$ of k elements and the rest of the system $X - X^k_j$ can be expressed in terms of mutual information: $MI(X^k_j, X - X^k_j) = H(X^k_j) + H(X - X^k_j) - H(X)$*

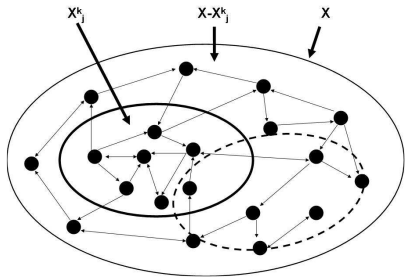

Curvature R near 1 indicates a subs-net which is as interactive with the rest of the system as they are within their subset. On the other hand, a R much higher than 1 will indicates the presence of a *functional cluster*- a subset of elements which are strongly interactive among themselves but only weakly interactive with the rest of the system. Function cluster is a manifestation of specialized region that involve in generation of high-level function. Geometrically, it will emerge as a horn area on information manifold.

Evolution of causal interactions among functional clusters is described by Ricci tensor $R_{ij}$ which is geometric analog to the concepts of *effective information* and *information integration* [9]. Effective information $EI(X^k_j \rightarrow X - X^k_j)$ between sub-net $X^k_j$ and $X - X^k_j$ can be seen as an amount of informational entropy that $X - X^k_j$ shares with $X^k_j$ due to causal effects of $X^k_j$ on $X - X^k_j$.

## Geometric Flow as Delocalization and Functionality Conservation

Summarize previous chapters, we propose model where an evolution of the system and the *architecture transition can be seen as a geometrical flow of information entropy on some informational manifold*. Results [10] demonstrates that Ricci Flow can be considered as renormalization semi-group that distribute informational curvature over the manifold but keep invariant $R = R_{min} * V^{2/3}$ where R-curvature and V-volume on information manifold. Region with strong curvature interpreted as sub-system with high information integration and recursive complexity. Thus, we proposed Ricci flow as a process of delocalization that provides distributed representation under architecture transition. Based on Perelman works [10] for the solutions to the Ricci flow (d/dt $g_{ij}(t) = -2R_{ij}$) the evolution equation for the scale curvature on Riemann manifold:

$$d/dt\ R = \Delta R = 2\ |Ric|^2 = \Delta R + 2/3\ R^2 + 2|Ric^o|^2$$

It implies the estimate $R^t_{min} > -3/(2*(t+1/4))$ where the larger t-scalar parameter then the larger is the distance scale and smaller is the energy scale.

The evolution equation for the volume is $d/dt\ V < R_{min}V$. Take R and V asymptotic at large t, we have $R(t)V(t)^{-2/3} \sim -3/2$. Thus, we have Ricci flow as a process of delocalization where V-growth when R-decrease that provides distributed representation under architecture transition.

*Figure 4. Demonstrate delocalization of functional cluster-dynamic core due to geometry flow*

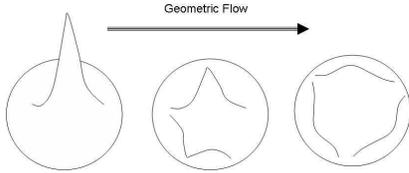

Another evolutional equation that can be introduced as a candidate for delocalization is Calabi flow which, unlike the Ricci flow, is only defined on Kahler manifolds with complex coordinates ($z_i$, $z'_j$):

$\partial g_{ij}/\partial t = \partial R/\partial z_i \partial z'_j$. A geometrical context for the Calabi flow represents spherical waves of informational metrics.

*Figure 5. Demonstrate distributed representation of a Dynamic Core on 3D informational manifold as a set of concentric conics representing individual points. Each point of the Dynamic Core is a state of an information system and the object itself is an assembly of causally related information objects.*

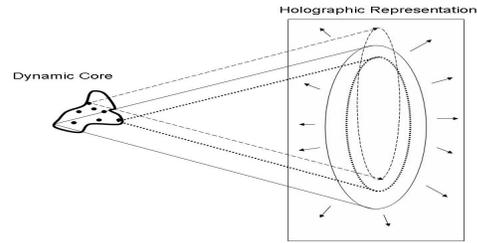

Define informational metric as $g_{ij} = 2(exp(\Phi(z,z',t)))$ – exponent of a scalar then two flows assume the same form Ricci: $\partial \Phi/\partial t = \Delta \Phi$ and Calabi: $\partial \Phi/\partial t = -\Delta\Delta\Phi$

## Coherent structure Pfaff Dimension and Topological Torsion as a self poetic functionality

We also investigate topological evolution of informational manifold. Dynamic core is viewed as a coherent structure such as deformable connection domain on information manifolds with certain similar topological properties. We consider topological properties of such space that stay the same under continuous transformations. For the scalar function X evolves in time with some velocity VX, following the method of Cartan [11], a certain amount of topological information can be obtained by the construction of the Pfaff sequences based on the 1-form of Action, $A = ds = VX_\mu(x)dx^\mu$ a differential constructed from the unit tangent information integration velocity field VX. We claim that emergent states are coherent topological structures:

*Topological Action (energy) A*
*Topological Vorticity (rotation) F=dA*
*Topological Torsion (entropy) H=A ∧ dA*
*Topological Parity (Dynamic Core) K = dA ∧ dA*

The rank of largest non-zero element of the above sequence gives Pfaff dimension of an information manifold. It gives us the minimal number M of functions required to determine the topological properties of the given form in a pre-geometric variety of dimensions N. We require at least dimension 4 to accommodate complex systems with dynamic cores.

It may be demonstrated from deRham theorem and Brouwer theorem [11] that the odd dimensional set (1,3,..) may undergo topological evolution but even dimensions remain invariant. It implies that Dynamic Core coherent topological structure once established through evolution of the Pfaff dimension from 3-to -4 then will remain invariant. Also, the Pfaff dimension is an invariant of a continuous deformation of the domain thus it is invariant under geometrical flow. We developed one particular example of such transition using Ising model [Appendix B].

**CONCLUSION**

We propose the information geometry to analyze recovery of lost functions in dynamic network systems. Also suggested, that an arbitrary functionality can be represented in terms of Conformal Yang-Mills field theory (CFT) as well as in terms of supergravity (and string) theory. Witten result [6] demonstrates precise correspondence between d-dimensional CFT and supergravity on the product of d+1-dimensional AdS space with a compact manifold. AdS model was assigned to local d+1-dim implementation of the function by specialized network. On the other hand, d-dim CFT was suggested for holographic (delocalized) implementation which represent artificial environment. Geometrical flow suggested as a mechanism of transition from one representation to another.

Such transition can conserves topological properties and as such preserve causal relations within the functional dynamic core. It is seen as a process of the lost function recovery.

We formulated recovery *constraint constrain*: delocalized d-dimensional dynamic core (DC) conserve functional properties only if exist d+1 compact representation of the DC that is precise correspondent to the original DC manifold.

We speculate about *delocalization* in biological neural networks that might explain the recovery of some functions lost during stroke or cancer. One interesting example would be the folk story in which tumor was removed from brain area related to speech. The patient ability to speak was lost due to destruction of the area but then ability to speak was recovered in several months. Our approach provides mathematical formalism for proposed explanation that other areas of the brain take a part in recovered of the lost function. Also, that might provide mathematical formalism for another documented case of Reverse Parkinson's in Two Patients which had Fetal-cell Transplants [12]. Finally, we would like to suggest that prosthesis or artificial tissue [13], which are able effectively implement specific entropy flow, will successfully adjust network system and recover lost mental functions.

**REFERENCES**

1. Brain prosthesis passes live tissue test. T. Berger, etc., New Scientist October 2004 (http://www.newscientist.com/article.ns?id=dn6574)
2. World's first brain prosthesis revealed T. Berger, etc., New Scientist March 2003 (http://www.newscientist.com/article.ns?id=dn3488)
3. IEEE ENGINEERING IN MEDICINE AND BIOLOGY MAGAZINE. SEPTEMBER/OCTOBER 2005. **Restoring Lost**. **Cognitive Function**. Hippocampal-Cortical Neural Prostheses. Berger, T.W. Ahuja, A. Courellis, S.H. Deadwyler, S.A. Erinjippurath, G. Gerhardt, G.A. Gholmieh, G. Granacki.
4. Astonishing Hypothesis: The Scientific Search for the Soul. Francis Crick.
5. Semantically Based Data Integration Environment for Biomedical Research. Astakhov V, A Gupta, J Grethe, E Ross, D Little, A Yilmaz, M Martone, X Qian, S Santini, M Ellisman *Proceedings of the 19th IEEE International Symposium on Computer-Based Medical Systems* 2006
6. Anti De Sitter Space and Holography. Edward Witten (http://arxiv.org/abs/hep-th/9802150)
7. http://en.wikipedia.org/wiki/Information_geometry , http://en.wikipedia.org/wiki/Curvature_tensor
8. Theories and measures of consciousness: An extended fraimwork. Gerald Edelman,… 2006, 10.1073/pnas.0604347103
9. Measuring information integration. Giulio Tononi and Olaf Sporns *BMC Neuroscience* 2003, **4:**31
10. Perelman, Grisha (November 11, 2002). "The entropy formula for the Ricci flow and its geometric applications". arXiv:math.DG/0211159
11. Kiehn, R. M., (1990a) "Topological Torsion, Pfaff Dimension and Coherent Structures";, in: H. K. Moffatt and T. S. Tsinober eds, Topological Fluid Mechanics, (Cambridge University Press), 449-458 http://www22.pair.com/csdc/pd2/pd2fre1.htm
12. Fetal-cell Transplants Reverse Parkinson's in Two Patients. NEUROLOGY, News from Harvard Medical, Dental, and Public Health School http://focus.hms.harvard.edu/2005/Jun10_2005/neurology.shtml
13. Tissue Engineering. Volume 13, Number 2, pp231-433
14. arXiv:quantum-ph/0507029 v2 6Jul2005 None local Hamiltonians and Information Conservation Law Jian-Ming. Xiang-Fa Zhou, Zheng-Wei Zhou, and Guang-Can Guo Compression
15. **Matrix models** as non-local hidden variables theories. Authors: **Lee Smolin** Comments: 25 pages, latex, no figures Subj-class: High Energy Physics - Theory; **...**arxiv.org/abs/hep-th/0201031

## [Appendix A] Mediator

The example with "shortest path" can be extended to the information integration system - *mediator*. Usually, a *mediator* is a server that can communicate with remote data sources {D1, D2,..} and perform distributed queries. One of the key functional components of a mediator system is a *planner* that creates order-of-execution commands {Q7,Q2,Q3, ...} that should be sent to remote sources {D7, D2, D3, ...}. The planning algorithm can be quite complicated because it tries to determine which source should be queried first (D7) and what intermediate information should be send to subsequent sources {D2, D3, ...}. But such *planner* functionality also can implicitly emerge in distributed system of nodes with a few simple rules:
1) Starting node at A try to execute any of {Q}
2) Broadcast the result and all {Q} to all neighbors
3) A neighbor tries to execute any of {Q} and get a result, then mark message "passed me" (if the neighbor get message that it already passed then it drop the message)
4) Repeat wave of broadcasts from neighbors until all {Q} have been executed
5) The node executed the last sub-query then returns results to A

As we can see, planning functionality will be implicitly performed as result of broadcast wave ropagations. Obviously such system could consume considerable bandwidth and may be not feasible from an engineering point of view, but this example gives us a sense of the *problem of multiple implementations* for pre-defined functionality, conservation of information dynamics and emergence of functionality.

## [Appendix B] Matrix theory and Ising mode for Information Geometry

Ricci flow was proposed as an information entropy flow that conserves dynamic core functionality by providing holographic representation for the initially localized DC. Also we proposed a toy model for holographic representation which is based of Smolin [15] matrix model with hidden variables.

Any holographic representation sub-system $X_a^b$ can be seen as a matrix N*N –of N of informational objects, where diagonal "ii" represent holographic set of projected eclipses and internal information integration for "i" if it has internal structure. And off-diagonal element "ij" represent effective information between elements i and j.

Interactions among sub-systems can be expressed by an Action $S \sim m \int dt\ Tr\{X'^2_a + \omega^2 [X_a\ X_b][X^a\ X^b]\}$ where $X_a$ is N*N matrix that can be represented X=D+Q as sum of diagonal D=diag($d_1,d_2,..$) and none-diagonal pieces. Then action can we wrote as $S \sim m \int dt\ (L^D + L^Q + L^{int})$

Nelson's stochastic formulation of quantum theory emerges naturally as a description of statistical behavior of the eigenvalues with interaction potential of interaction between diagonal and none-diagonal elements:

$$L^{int} = U(D,Q)$$
$$L^D = m \sum D'^2_a$$
$$L^Q = m\{\sum Q'^2_a + \omega^2 [Q_a\ Q_b][Q^a\ Q^b]\}$$

$L^{int} = 2m\omega^2 \sum \{-(d_i-d_j)^{2a}\ Q^2_b - (d_i-d_j)_a\ (d_i-d_j)_b\ Q^a\ Q^b - 2(d_i-d_j)\ ^aQ^b\ [Q_a\ Q_b]\}$ based on this potential the classical equation of motion can be wrote how each matrix element moves in an effective potential created by the average motion of other elements. We assume that statistical averages satisfy (Gaussians processes) relations consistent with the symmetry of the theory. That gives use Brownian movements in potential:

$$<U> = m\Omega_Q^2/2\ Q_{ija}\ Q^{ija} + m\Omega_d^2/2\ (d_a^i-d_a^j)^2$$
$$\text{where}\ \Omega_Q^2 = 4(d-1)\omega^2\ [(N-1)\ q^2+2\ r^2]$$
$$\Omega_d^2 = 4(d-1)\omega^2\ q^2$$

The Q system is in distribution. Variation principle for Matrix Model can be reformulated for eigenvalues. $\lambda = d + \sum Q^{ij}_a\ Q^{ji}_a/(d_i-d_j)_a + \ldots$ . And Diffusion constant for the eigenvalues is $\nu = (\Delta d)^2/\Delta t$

Now we would like to show that neural-network can emulate quantum mechanical system at normal temperature. We are going to make one assumption that $T/(8(d-1)m\ \omega^2) = t/N^p$ and $\hbar = m\ \nu$, then we can define wave function: $\Psi = \sqrt{\rho} \exp(S/\hbar)$, where $\rho$ - probability density $= 1/Z \exp(-H(Q)/T)$ and Hamiltonian $H(Q) = m\{\sum Q'^2_a - \omega^2 [Q_a\ Q_b][Q^a\ Q^b]\}$.

That illustrate Lee Smolin result that variation principle in presence of Brownian motion equivalent to Schrödinger equation:

$$i \hbar \, d\Psi/dt = \{ -\hbar/2m \, d^2/d(\lambda)^2 + m\Omega_d^2/2 \sum (\lambda_a^i - \lambda_a^j)^2 + TN(N-1)/4 + NmC \} \Psi$$

The analysis provides analogy between neural network dynamics and effects of quantum theory. Now, we have Hamiltonian to try Ising approach for the information geometry approach. As we demonstrated in the paper, the geometric structure in which informational manifold is endowed leads to certain local invariants, one of the most important being the Ricci scalar curvature R. $R \sim \zeta^d$, where $\zeta$-is the correlation length, which is two-point function, and d- denotes the number of spatial dimensions of the model. The curvature also receives contribution from higher order correlations.

The scaling behavior of the curvature in the vicinity of the critical point provides a satisfying picture of how certain universal features of the near-critical regime are encoded in the Fisher-Rao geometry of the informational manifold.

We will call N*-critical size, above which the system is to a reasonable approximation 'thermodynamic'. When the size drop below N* the system often behaves in qualitatively different way. R is strictly positive in the thermodynamic regime and negative at none-thermodynamic regime.

We can start with the state of the system immersed in a large heat bath with fixed temperature T in thermal equilibrium. The Gibbs measure can be mapped to Hilbert space $P(x|\theta) = \exp(-\sum_{i=1:r} \theta_i H^i(x) - \ln Z(\theta)) = \psi_\theta(x)$ where $\{\theta_i\}$-represent r-dim sub-space S in the Hilbert space. The Fisher–Rao metric can be introduced on the maximum entropy manifold as : $g_{ij} = 4\int \partial_i \psi_\theta(x) \, \partial_j \psi_\theta(x) \, dx = \partial_i \partial_j \ln Z(\theta) = \partial_i \partial_j S(entropy)$, $\partial_i = \partial/\partial\theta^i$.

And an entropy $S(p|q) = \int p \ln(p/q) = S(p(\theta)|p(\theta+d\theta)) = 1/2 \, g_{ij} \, d\theta^i \, d\theta^j + ..$ is represented as Taylor series. Then manifold can be interpreted as a maximum entropy surface and consequently specific geodesics will correspond to equations of state for the system.

We take simple (Ising chain) source-to-source (H1) and source to net (H2) interaction then r=2 and $-\beta H = \beta \sum s_i s_j + h \sum s_i$, $\beta = 1/kT$, $s = \{+1, -1\}$, h-network "field" then Scalar curvature

$R = -1/(2 \det(g)) * \det\{$
$\partial_1^2 \ln Z \quad \partial_1 \partial_2 \ln Z \quad \partial_2^2 \ln Z$
$\partial_1^3 \ln Z \quad \partial_1^2 \partial_2 \ln Z \quad \partial_1 \partial_1^2 \ln Z$
$\partial_1^2 \partial_2 \ln Z \quad \partial_1 \partial_1^2 \ln Z \quad \partial_2^3 \ln Z$
$\}$

will act as an indicator of finite size effects. Differentiate $N^{-1}$ in Z we have
$g_{ij} = 1/N \, \partial_i \partial_j \{N \beta + \ln[(\cosh h + \eta)^N + (\cosh h - \eta)^N]\}$, where $\eta = (\sinh^2 h + \exp(-4\beta))^{1/2}$

Thus the size of dynamic core depends on temperature as T~size.
Through Ricci flow R<0 (localized DC) can transit to R>0 delocalized implementation.
if N->∞ then we have thermodynamic curve $R = 1 + \eta^{-1} \cosh h > 0$.

Figure 5. Process of Holographic representation can be seen as Geometric flow that smooth and delocalize curvature. Region with strong curvature interpreted as sub-system with high information integration and recursive complexity

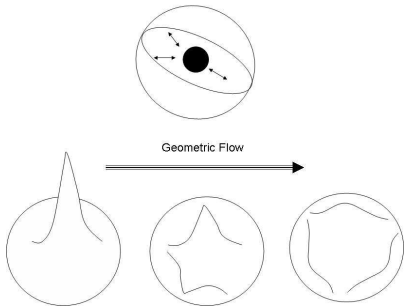

Geometric Flow

Now, lets consider just two sources with interaction described by a Hamiltonian: $H = \sum C_i * \sigma_i * \tau_i$.
The prescribed state is denoted by the density matrix $\rho = |\psi\rangle\langle\psi|$, where $|\psi\rangle = \cos(\theta/2)|0\rangle + \exp(i\varphi)\sin(\theta/2)|1\rangle$ is superposition of active $|1\rangle$ and $|0\rangle$ none-active state. As we can see an initial state of the system of two sources is $|\rho(0)|_{4\times 4}$. The information encoded in the system at time t is characterized by the fidelity $F_i(t) = \langle \psi | \rho^i(t) | \psi \rangle$, where $\rho^i(t) = \text{Tr} \, \rho(t)$-except "i".

Let's introduce two qualities:
$$CF_i(t) = \cos^2(\theta/2)\, \rho^i_{00}(t) + \sin^2(\theta/2)\, \rho^i_{11}(t)$$
$$QF_i(t) = \mathrm{Re}[\exp(-i\varphi) * \sin(\theta)\, \rho^i_{10}(t)]$$

And rewrite Fidelity as $F_i(t) = CF_i(t) + QF_i(t)$, where $\rho(t) = \exp(-iHt)*\rho(0)*\exp(iHt)$.

By straightforward calculations [14], we found that if $C_i = C_j$ (uniformity of information integration among all sources) for any i and j then $\sum CF_i(t)$ and $\sum QF_i(t)$ are both invariable. That mean total $F(t) = \sum F_i(t)$ is invariable too.

Due to the interactions between parts of the system, the states of these parts changes with time, and the information is expanded between them. But the total Fidelity-information is conserved. This is something like Energy Conservation Law for information systems. This law leads to interesting phenomena when information can partially concentrated spontaneously in one part of the whole system due to oscillation part of $F_i(t)$. Such concentration and following dilution in large-scale system with thermal bath can be seen as localization and de-localization of the Dynamic Core.

The theorem [14] also gives us a mechanism to construct informational systems (universes) with conservative dynamic core. Such universes should have "conservative Hamiltonian" that will preserve total information integration. Hamiltonian should satisfy condition $[H, \sum_{i=1}^{n} C_i] = 0$, where $C_i = 1/(2^{n-1})\, I_1 * I_2 * \ldots \rho^0_i * I_{i+1} * I_{n+2} * \ldots I_n$, with $\rho^0_i$ denotes the reduced density matrix of the i-th source and $I_i$ denotes reduced density matrix of other sources. This is easily can be proven by showing that $\partial F(t)/\partial t = 0$ is equal to $[H, \sum_{i=1}^{n} C_i] = 0$.